\documentclass[reprint,nofootinbib,amsmath,amssymb,aps]{revtex4-2}

\usepackage{graphicx} 

\usepackage{amssymb}
\usepackage{amsmath}
\usepackage{subfigure}
\usepackage{enumitem}
\usepackage{color}
\usepackage{hyperref}
\usepackage{float}
\usepackage{here}
\usepackage{physics}

\newcommand{\hGpc}{$ \, h^{-1} \rm Gpc$}
\newcommand{\hMpc}{$ \, h^{-1}  \rm Mpc$}
\newcommand{\hMpcinv}{$ \, h \, \rm Mpc^{-1}$}
\newcommand{\hMpccube}{$(h^{-1} \rm Mpc)^3$}

\newcommand{\hkpc}{$ \, h^{-1} \rm kpc$}

\newcommand{\hMsun}{$\, h^{-1} \rm M_\odot$}

\newcommand{\kms}{$\, \rm km \, s^{-1}$}
\newcommand{\lsim}{\mbox{${\,\hbox{\hbox{$ < $}\kern -0.8em \lower 1.0ex\hbox{$\sim$}}\,}$}}
\newcommand{\gsim}{\mbox{${\,\hbox{\hbox{$ > $}\kern -0.8em \lower 1.0ex\hbox{$\sim$}}\,}$}}
\newcommand{\lcdm}{Planck18}
\newcommand{\lcdmdde}{Planck18$+$DDE}
\newcommand{\desidde}{DESIY1$+$DDE}
\newcommand{\desi}{DESIY1}
\newcommand{\urldde}{\url{https://skun.iaa.csic.es/SUsimulations/DDE/}}
\newcommand{\SU}{{Skies \& Universes}}

\newcommand\apjs{{\it Astrophys. J.Supp.}}
\newcommand\aap{{\it Astron. Astrophys.}}
\newcommand\mnras{{\it Mon. Not. R. Astron. Soc.}}
\newcommand\pasj{{\it Publ. of the Astron. Society of Japan}}
\newcommand\na{{\it New Astronomy}}
\newcommand\jcap{J. Cosmology Astropart. Phys}

\begin{document}
\title{Evolution of clustering in cosmological models with time-varying  dark energy}

\author{Tomoaki Ishiyama}
\email[e-mail:]{ishiyama@chiba-u.jp}
\affiliation{Digital Transformation Enhancement Council, Chiba University, 1-33, Yayoi-cho, Inage-ku, Chiba, 263-8522, Japan}
\author{Francisco Prada}
\affiliation{Instituto de Astrof\'isica de Andaluc\'ia (CSIC), Glorieta de la Astronom\'ia, E-18080 Granada, Spain}
\author{Anatoly A. Klypin}
\affiliation{Astronomy Department, New Mexico State University, Las Cruces, NM, USA}
\affiliation{Department of Astronomy, University of Virginia, Charlettesville, VA, USA}

\preprint{}
\date{\today}

\begin{abstract}
Observations favor cosmological models with a time-varying dark energy component. But how does dynamical dark energy (DDE) influence the growth of structure in an expanding Universe?
We investigate this question using high-resolution $N$-body simulations based on a DDE cosmology constrained by first-year DESI data (\desidde), characterized by a 4\% lower Hubble constant ($H_0$) and 10\% higher matter density ($\Omega_0$) than the Planck-2018 $\Lambda$CDM model.
We examine the impact on the matter power spectrum, halo abundances, clustering, and Baryonic Acoustic Oscillations (BAO). 
We find that \desidde\ exhibits a 10\% excess in power at small scales and a 15\% suppression at large scales, driven primarily by its higher $\Omega_0$.
This trend is reflected in the halo mass function: \desidde\ predicts up to 70\% more massive halos at $z = 2$ and a 40\% excess at $z = 0.3$. 
Clustering analysis reveals a 3.71\% shift of the BAO peak towards smaller scales in \desidde, consistent with its reduced sound horizon compared to \lcdm. Measurements of the BAO dilation parameter $\alpha$, using halo samples with DESI-like tracer number densities across $0 < z < 1.5$, agree with the expected \desidde-to-\lcdm\ sound horizon ratio. After accounting for cosmology-dependent distances, the simulation-based observational dilation parameter closely matches DESI Y1 data. 
We find that the impact of DDE is severely limited by current observational constraints, which strongly favor cosmological models -- whether including DDE or not -- with a tightly constrained parameter $\Omega_0h^2\approx 0.143$, within 
1-2\% uncertainty. Indeed, our results demonstrate that variations in cosmological parameters, particularly $\Omega_0$, have a greater influence on structure formation than the DDE component alone. 
\end{abstract}

\maketitle

\section{Introduction}\label{sec:intro}

Measurements of the Baryon Acoustic Oscillations (BAO) feature in the clustering of galaxies from the
first- and third-year data of the Dark Energy Spectroscopic Instrument
(DESI Y1 and Y3~\cite{DESI2024,DESIY3}), combined with the Cosmic Microwave
Background (CMB) anisotropies from the Planck
satellite (\lcdm~\cite{Planck2020}), indicate a preference for a time-varying Dynamic Dark Energy (DDE) equation of state, with a deviation from $\Lambda$CDM at the $\sim$3.1$\sigma$ level.
Including type Ia supernovae (SN Ia) data alongside with \desi\ BAO and CMB measurements further increases the statistical significance of DDE signal, reaching between $\sim$2.8$\sigma$ and $\sim$4.2$\sigma$, depending on the specific SN Ia dataset used (see \cite{DESIY3}).   

Several DDE models have been proposed in the literature~\cite{Ratra1988,Chevallier2001, Linder2003,JassalDDE,BarbozaDDE,Ray2011,Yin2022}. Among them, the
Chevallier-Polarski-Linder (CPL) parametrization~\cite{Chevallier2001,Linder2003} is one of the most widely used formulations of a time-varying dark energy equation of state, and is frequently adopted in the analysis of observational data. In this model, the dark energy equation of state parameter $w$ evolves with the scale factor $a$ as:
\begin{eqnarray}
w(a) = w_0 + w_a\qty(1-a), 
\end{eqnarray}
where $w_0$ and $w_a$ are free parameters. 

Notably, the values of the matter density $\Omega_0$
and Hubble constant ($H_0=100\,h\,$km/sec/Mpc), derived when inlcuding DESI data, differ significantly from those inferred from Planck-only constraints. The
\desi\ data suggests a $\sim10\%$ higher matter density and a $\sim4\%$ lower Hubble constant, with best-fit DDE parameters
$w_0=-0.45$ and $w_a=-1.79$. The resulting cosmological parameters, summarized in
Table~\ref{tab:model}, define the \desidde\ cosmology used throughout this work. For comparison, we also consider the baseline \lcdm\ model based on Planck-2018 results~\cite{Planck2020}, as well as a hybrid model \lcdmdde\ that adopts Planck-2018 values for standard cosmological parameters but uses the $w_0$ and $w_a$ parameters from \desidde.
Table~\ref{tab:model} also gives the corresponding values of the baryon drag epoch $z_{\rm d}$ and the comoving sound horizon scale $r_{\rm d}$. 
The difference of these parameters mainly comes from that of cosmological parameters among models.
These differences among the three models can have a substantial impact on various large-scale structure (LSS) observables, including the power spectrum, the abundance of galaxies (dark mater halos), and their clustering properties. 

\begin{figure*}
\centering 
\includegraphics[width=0.32\textwidth]{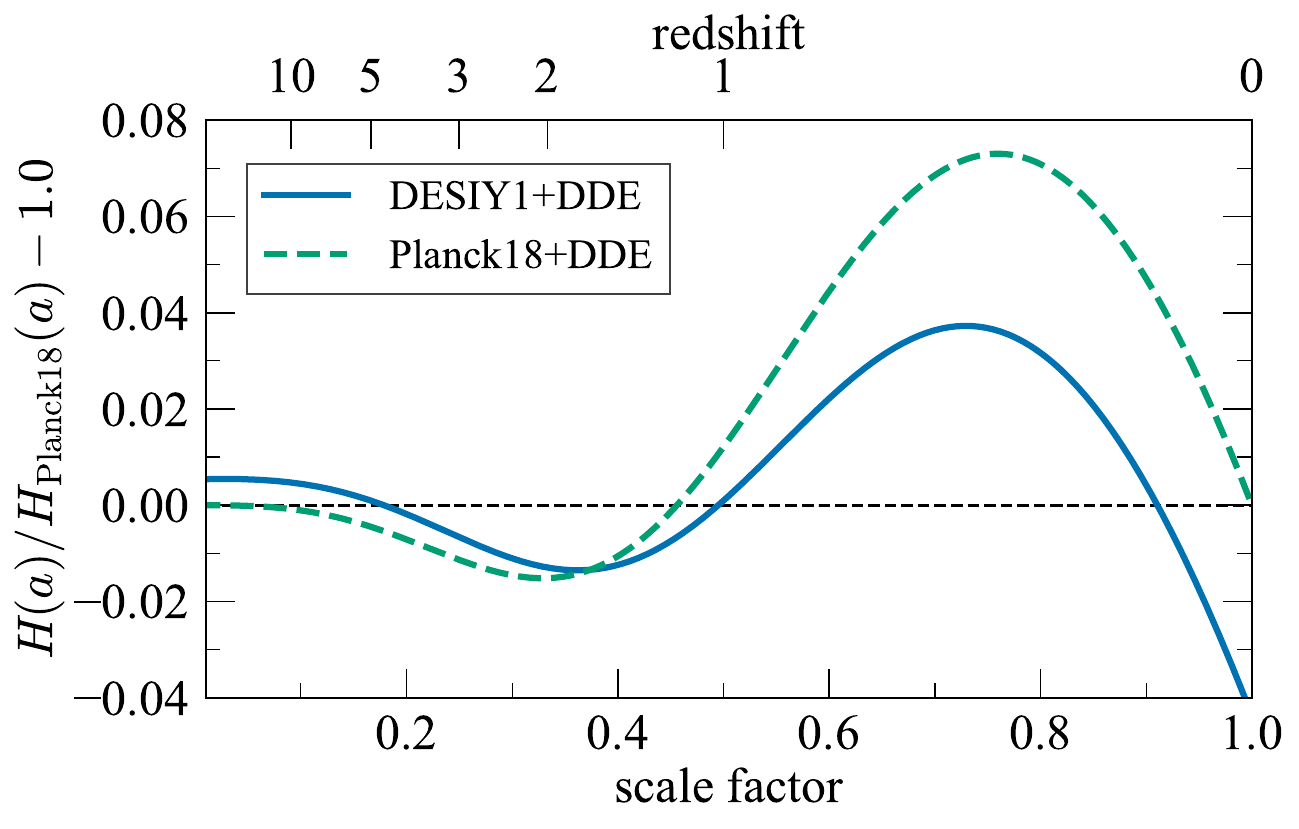}
\includegraphics[width=0.32\textwidth]{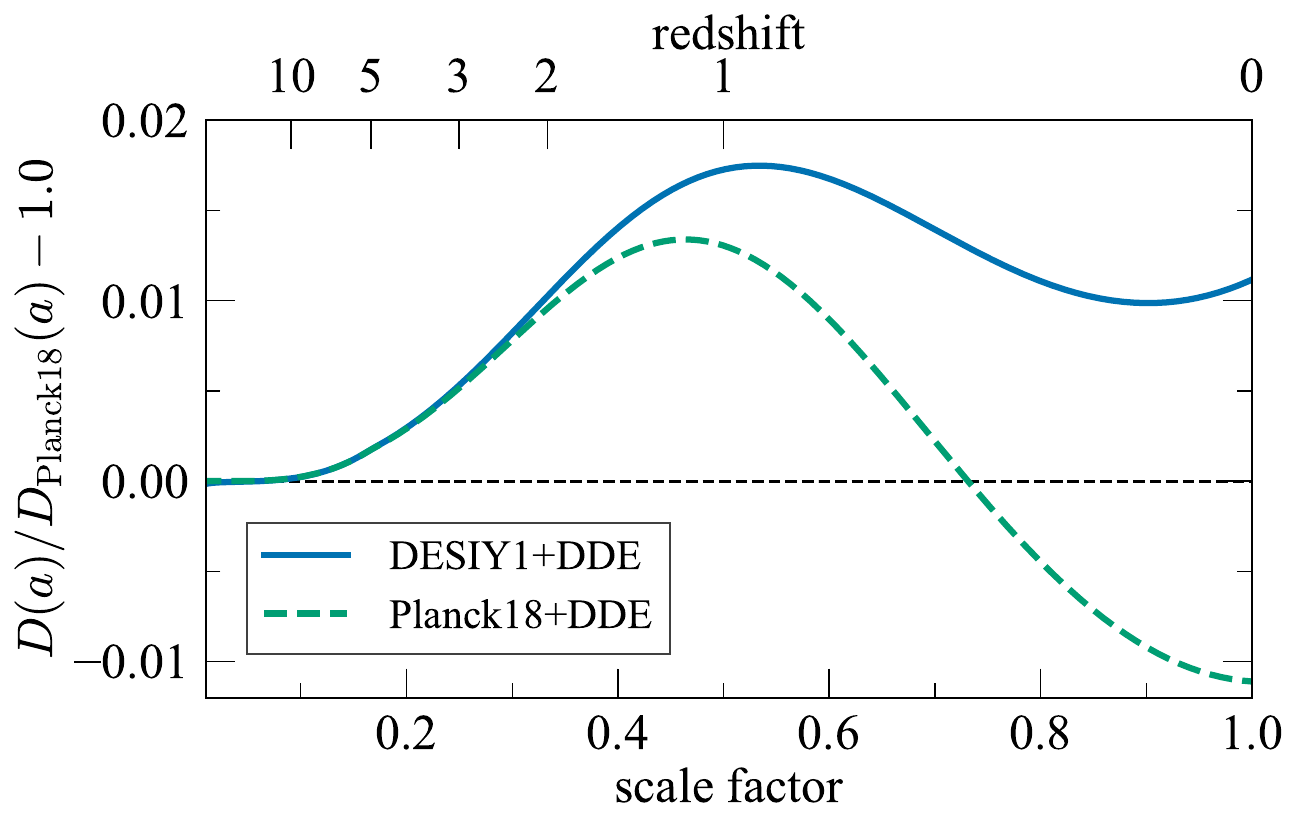}
\includegraphics[width=0.32\textwidth]{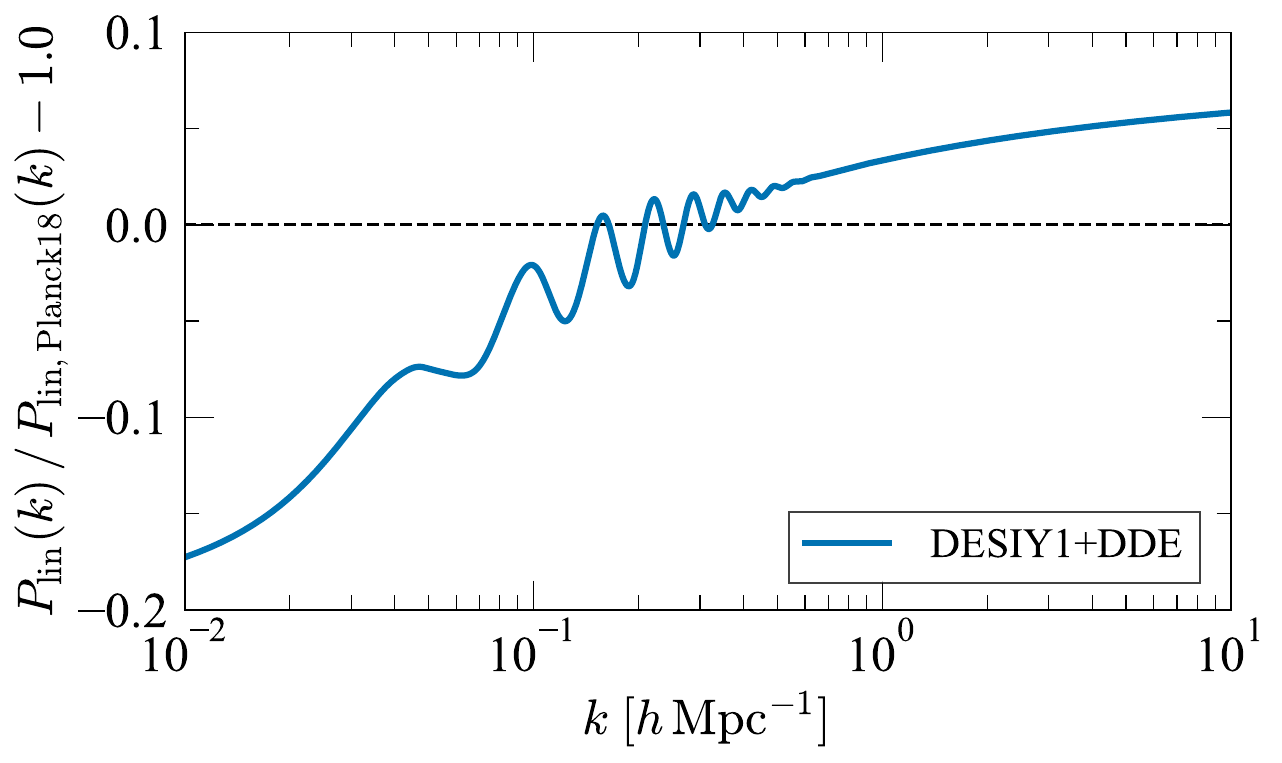}
\caption{Evolution of the Hubble parameter (left), linear growth factor (middle), and the linear matter power spectrum at $z=24$ (right) for the cosmological models indicated in the labels, shown relative to the \lcdm\ model. }
\label{fig:model} 
\end{figure*}

Cosmological simulations are powerful tools to studying non-linear structure growth and for accurately estimating key statistical measures. A number of studies have employed $N$-body simulations to explore DDE models ~\cite[e.g.,][]{Klypin2003,CasariniDDE,Alimi2010,Pfeifer2020,Beltz-Mohrmann2025}.  However, the volumes of recent galaxy surveys are extremely large, leading to relatively small statistical uncertainties. To match this precision, high-resolution cosmological simulations based on DDE must cover similarly large volumes. Such large-volume simulations are not that common for DDE, in sharp contrast to the extensive suite of concordance $\Lambda$CDM simulations~\cite[e.g.,][]{Potter2017,Heitmann2019,Ishiyama2021,Wang2022,Frontiere2022}. 

\citet{Beltz-Mohrmann2025} performed two large $N$-body simulations of DDE models, each with a box size of 1\hGpc. Their results on the nonlinear dark matter power spectra and halo mass functions are qualitatively consistent with our findings. In this regard, their results and ours are complimentary. However, \citet{Beltz-Mohrmann2025} did not study the clustering of dark matter halos, nor did they analyze fluctuations in the BAO domain - an essential aspect for interpreting DESI observations and one of the main focuses of our work. Such an analysis requires either a very large computational box or many realizations to overcome the significant cosmic variance at BAO scales. Our simulations enable this analysis, as their total volume is eight times larger than that of \citet{Beltz-Mohrmann2025}.

In this work, we perform large
cosmological $N$-body simulations using cosmological parameters that include  DDE parameters constrained by \desi\ data, and investigate the impact of DDE on various LSS
statistics.  This paper is organized as follows. Dynamical Dark Energy models are introduced in Sec.~\ref{sec:DDE}. In
Sec.~\ref{sec:sim}, we describe the cosmological
simulations in detail. Results are presented in
Sec.~\ref{sec:result}, followed by a discussion in  Sec.~\ref{sec:discussion}. Our main findings are summarized in Sec.~\ref{sec:summary}.

\begin{table}[tbp]
\centering
\caption{Estimates of
cosmological parameters with 68\% credible intervals from the Planck (\lcdm)~\cite{Planck2020} and DESI (\desidde)~\cite{DESI2024} experiments are shown.  
Here, $z_{\rm d}$ denotes the baryon drag epoch, and $r_{\rm d}$ is the comoving sound horizon scale.
In addition to these two models,  
we also consider a third model in this paper, denoted as (\lcdmdde), which adopts the same cosmological parameter as \lcdm, but includes the effects of dynamical dark energy.~\footnote{The inflation scenario is the same across the models.}
}
\begin{tabular}{lccc}
\hline
Parameter & \lcdm\ & \lcdmdde & \desidde \\
\hline
$\Omega_0$ & $0.3111 \pm 0.0056$ & 0.3111 & $0.3440^{+0.032}_{-0.027}$ \\
$h$ & $0.6766 \pm 0.0042$ & 0.6766 & $0.6470^{+2.2}_{-3.3}$ \\
$w_0$ & $-1.0$ & $-0.45$ & $-0.45^{+0.34}_{-0.21}$ \\
$w_a$ & $0.0$ & $-1.79$ & $-1.79^{+0.48}_{-1.0}$ \\
$z_{\rm d}$ & 1060.02 & 1060.02 & 1055.70 \\  
$r_{\rm d}$ [\hMpc] & 99.61 & 99.61 & 96.05 \\
\hline
\end{tabular}
\label{tab:model}
\end{table}


\section{Dynamical Dark Energy cosmology}\label{sec:DDE}
The evolution of the Hubble constant in DDE model with CPL parametrization is given by 
\begin{eqnarray}
\qty[\frac{H(a)}{H_0}]^2 =  \frac{\Omega_0}{a^3} + \frac{\lambda_0}{a^{3\qty(1+w_a+w_0)}} \exp(-3w_a\qty(1-a)).
\label{eq:dde_hubble}
\end{eqnarray}
The left and middle panels in Figure~\ref{fig:model} show the evolution of the Hubble parameter and linear growth factor for the \desidde\ and \lcdmdde\ models, relative to those of the \lcdm\ model. Both quantities deviate from the \lcdm\ model by at most $\sim 4\%$. While these differences are relatively small, they are potentially measurable. Most of the deviations occur at low redshifts ($z<1$, or equivalently, scale factor $a>0.5$), whereas at higher redshifts the differences become negligible.

The main cosmological parameters of the three models studied in this work are listed in  Table~\ref{tab:model}. All other parameters are identical across the models: $\Omega_{\rm b}=0.048975$, $n_{\rm s}=0.9665$, and
$\sigma_8=0.8102$. It is important to note that power spectrum of density fluctuations  depends critically on a combination $\Omega_0h^2$, which determines the location of the peak in the power spectrum. The trio of parameters $\Omega_0h^2$, $n_{\rm s}$ and $\sigma_8$
nearly - but not entirely - fixes the shape and amplitude of the linear perturbation spectrum.
In all three models considered here, these three parameters are the same. What differes among the models is the baryons fraction $\Omega_{\rm b}/\Omega_0$, which influences the power spectrum at the $\sim 10\%$ level and is primarily responsible for the differences in the BAO features observed in the spectrum.

The right panel of Figure~\ref{fig:model} shows the differences in the linear power spectra between the \desidde\ and \lcdm\ models. The \lcdmdde\ model shares the same linear power spectrum as \lcdm, since they have identical early-Universe parameters. In contrast, the \desidde\ model exhibits notable deviations: an increase in power of $\sim 5\%$ at small scales $k>1$\hMpcinv\ and a decline of $\sim 20\%$ at large (Gpc) scales $k<0.01$\hMpcinv. These changes in the shape of the power spectrum are an indirect consequence of adopting a dynamical dark energy model. While DDE does not directly impact the physics of the early Universe that sets the linear power spectrum of fluctuations, adopting DDE parameters leads to different best-fit values of $H_0$, $\Omega_0$ and  $\Omega_{\rm baryon}$ that remain consistent with observational constraints. It is these altered parameters that drive the significant modifications observed in the power spectrum.
\begin{figure*}
\centering 
\includegraphics[width=0.46\textwidth]{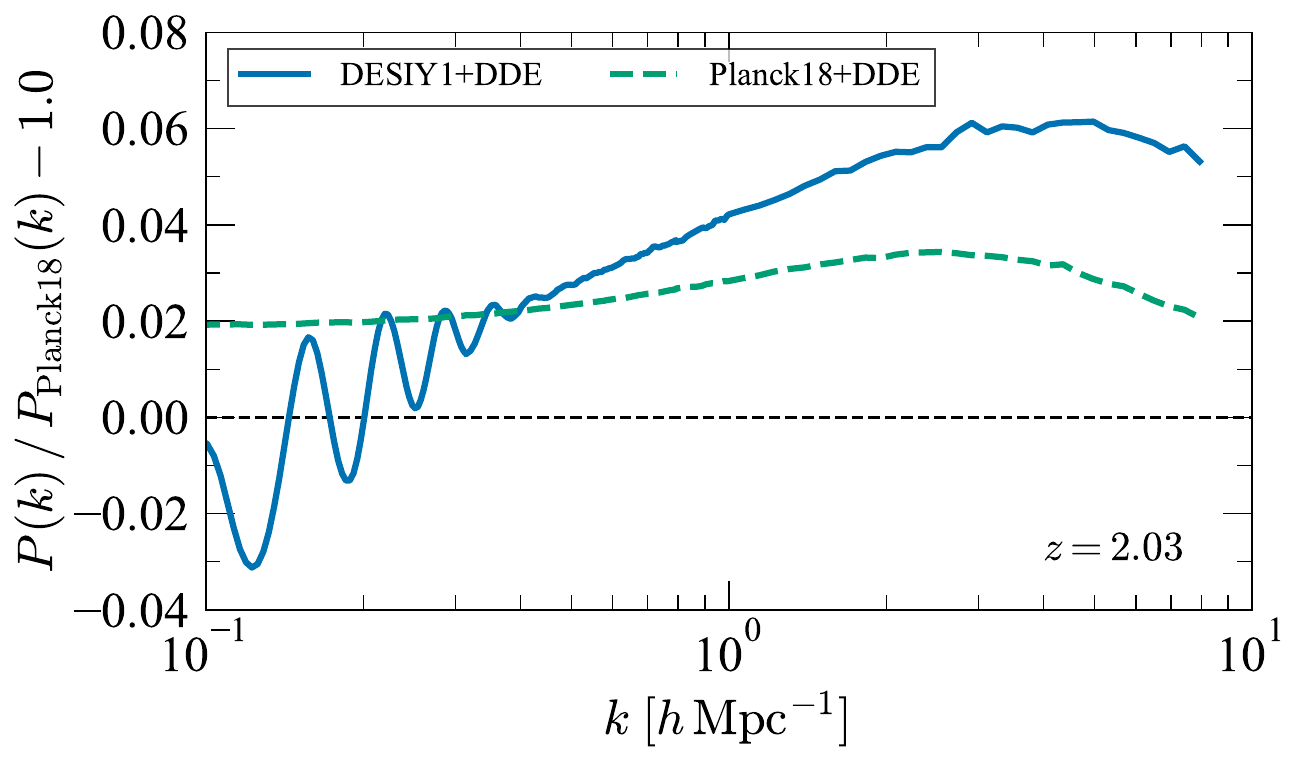}
\includegraphics[width=0.46\textwidth]{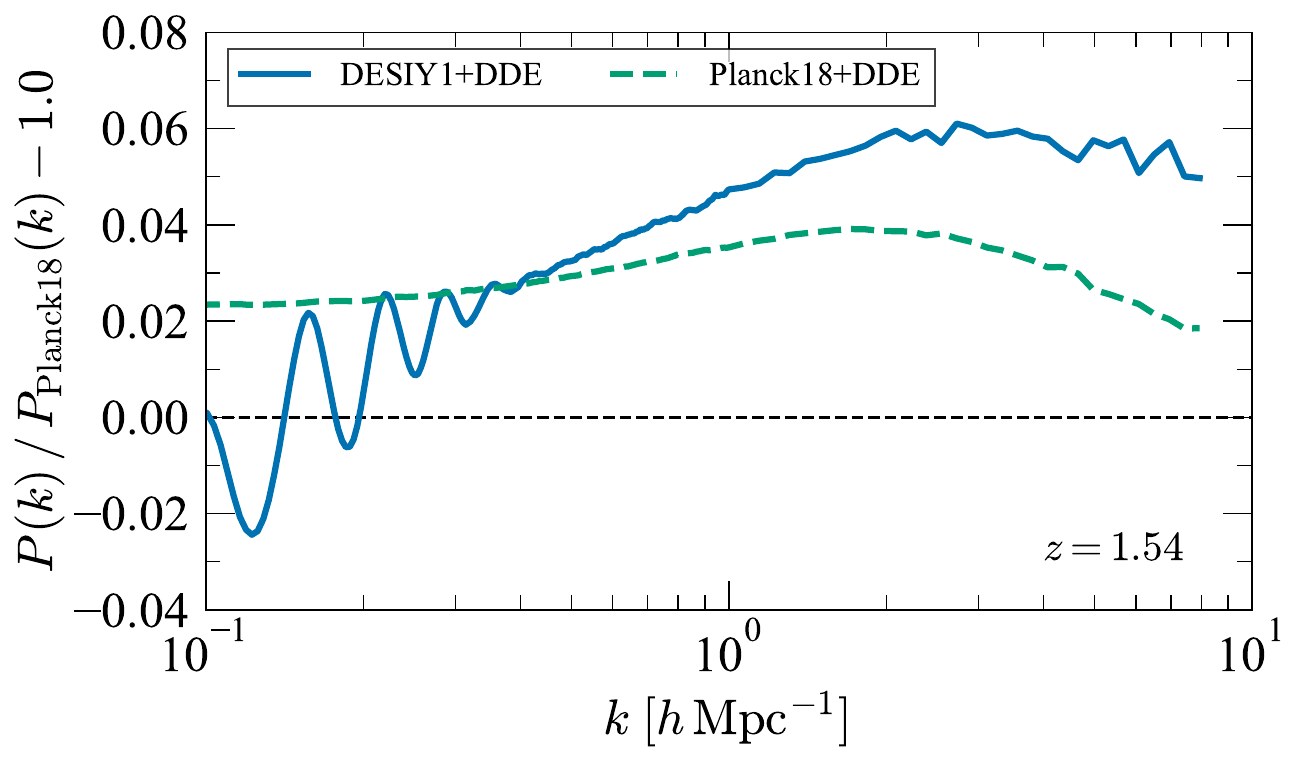}
\includegraphics[width=0.46\textwidth]{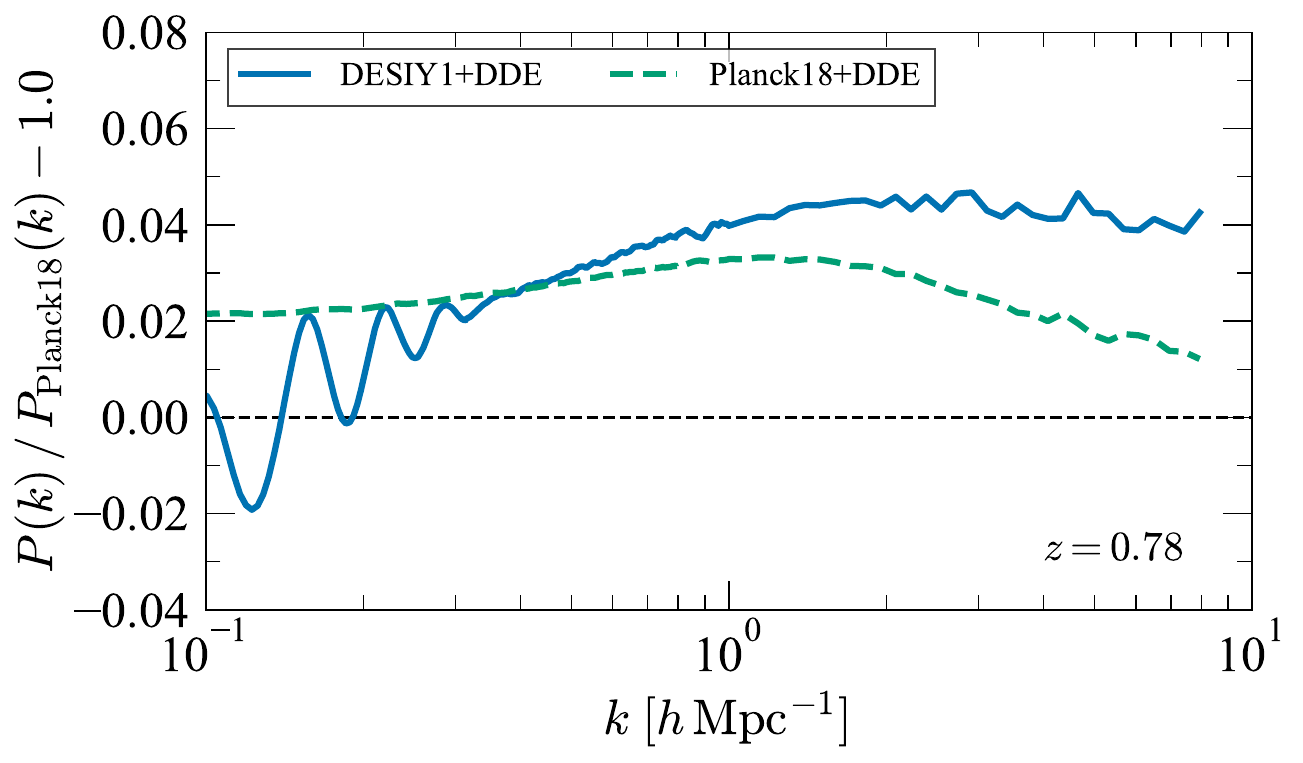}
\includegraphics[width=0.46\textwidth]{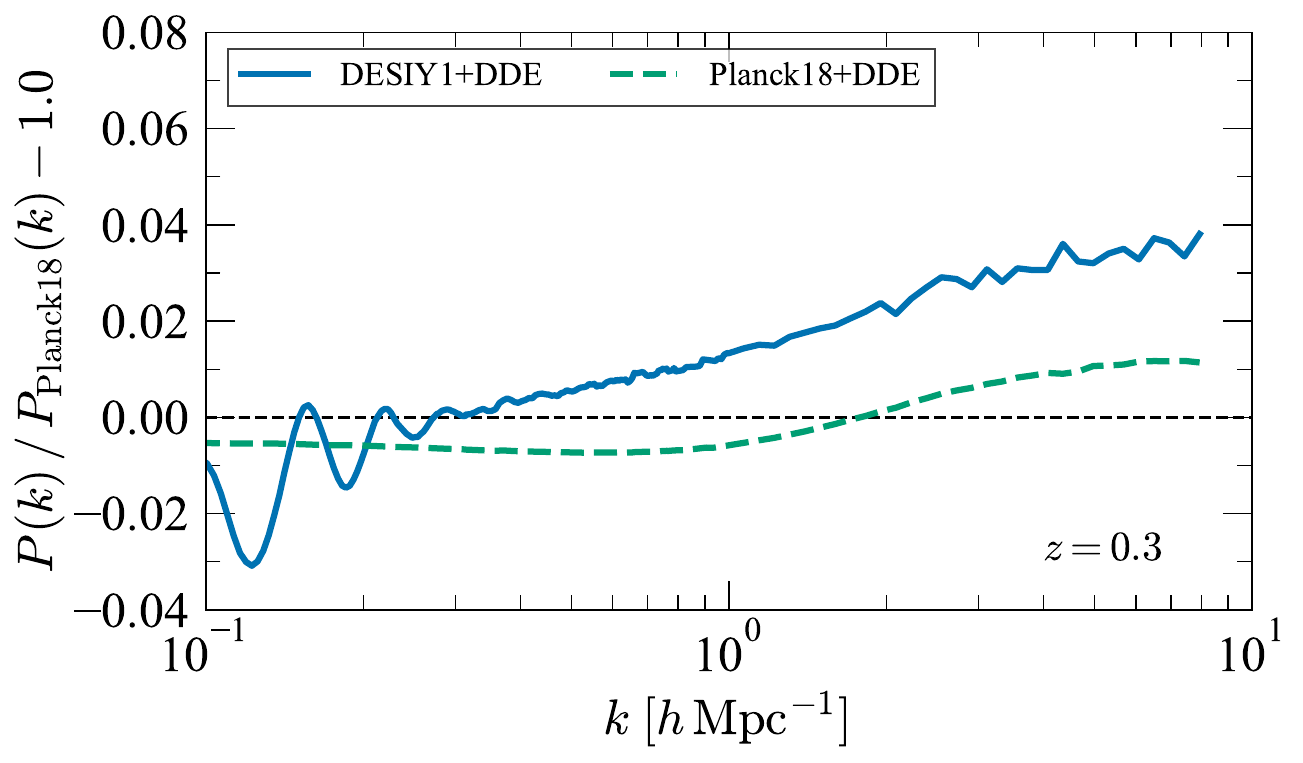}
\caption{Matter power spectra of the \desidde\ and \lcdmdde\ models relative to the \lcdm\ simulation at $z=2.03, 1.54, 0.78$, and 0.30.
The relative differences decrease toward lower redshifts, although the power spectra differ by less than 10\% at all redshifts.
The difference in cosmological parameters between the \lcdm\ and \desi-based models has a more significant impact on the power spectrum than the effect of dynamical dark energy.}
\label{fig:pk} 
\end{figure*}
The increase in power spectrum at large wavenumbers significantly impacts the evolution of the halo mass function and, consequently, galaxy formation. Due to the higher amplitude of fluctuations, halos collapse earlier,  leading to an overall increase in their abundance. This shift in amplitude affects massive halos more strongly than low-mass halos. In summary, the DDE models \desidde\ and \lcdmdde\ are expected to produce a higher number of halos (and hence galaxies) compared to the \lcdm\ model.

\section{Cosmological $N$-body simulations}\label{sec:sim}

We perform three cosmological $N$-body simulations, each with a  
box size of $L=2000$\hMpc\ and a particle number of $N = 4096^3$.
The particle mass resolution is $m_{\rm p} = 1.0
\times 10^{10}$\hMsun\ for the \lcdm\ and \lcdmdde\ simulations, and
$m_{\rm p} = 1.1 \times 10^{10}$\hMsun\ for the \desidde\ simulation.

Initial conditions were generated using the publicly available code,
\textsc{monofonIC}~\footnote{\url{https://bitbucket.org/ohahn/monofonic/}}~\cite{MONOFONIC:3LPT,MONOFONIC:BARYONS},
using third-order Lagrangian perturbation theory.  All three simulations share the same initial
phase and starting redshift ($z_{\rm ini}=24$), allowing us to ignore the impact of cosmic variance. The matter transfer functions were computed using the Python version of
\textsc{Camb}~\footnote{\url{https://camb.info/}}\cite{Lewis2000}.

We follow the non-linear gravitational evolution of dark matter using the massively parallel TreePM code
\textsc{GreeM}~\footnote{\url{http://hpc.imit.chiba-u.jp/~ishiymtm/greem/}}~\cite{Ishiyama2009b,Ishiyama2012,Ishiyama2022}, run
on the Fugaku supercomputer at the RIKEN Center for Computational Science. To incorporate the effects of dynamical dark energy, 
we implemented Eq~\eqref{eq:dde_hubble} into the \textsc{GreeM} code. The plummer
gravitational softening length was set to $\varepsilon=8.0$ \hkpc.  We used the
\textsc{Phantom-grape}~\footnote{\url{https://bitbucket.org/kohji/phantom-grape/src}}
code~\cite{Nitadori2006, Tanikawa2012, Tanikawa2013, Yoshikawa2018}
to accelerate the pairwise gravity force calculation.
Halos and subhalo identification was performed using \textsc{MPI-Rockstar}~\footnote{\url{https://github.com/Tomoaki-Ishiyama/mpi-rockstar/}}~\cite{Tokuue2024}, a hybrid MPI and OpenMP massively parallel code,
which is an extension of the original \textsc{Rockstar} phase space (sub)halo
finder~\footnote{\url{https://bitbucket.org/gfcstanford/rockstar/}}~\cite{Behroozi2013}.
The resulting HDF5 halo catalogs are publicly available on the \SU\ site at \urldde.

\begin{figure*}
\centering 
\includegraphics[width=0.46\textwidth]{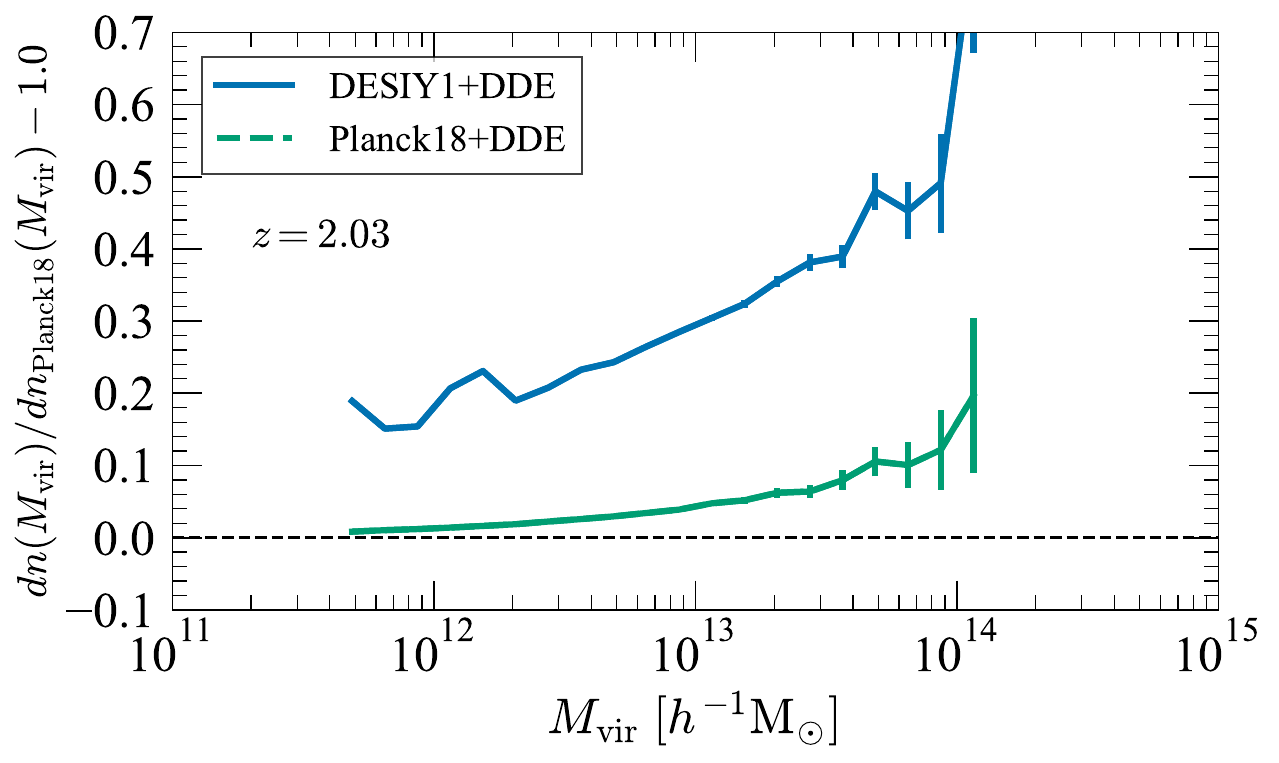}
\includegraphics[width=0.46\textwidth]{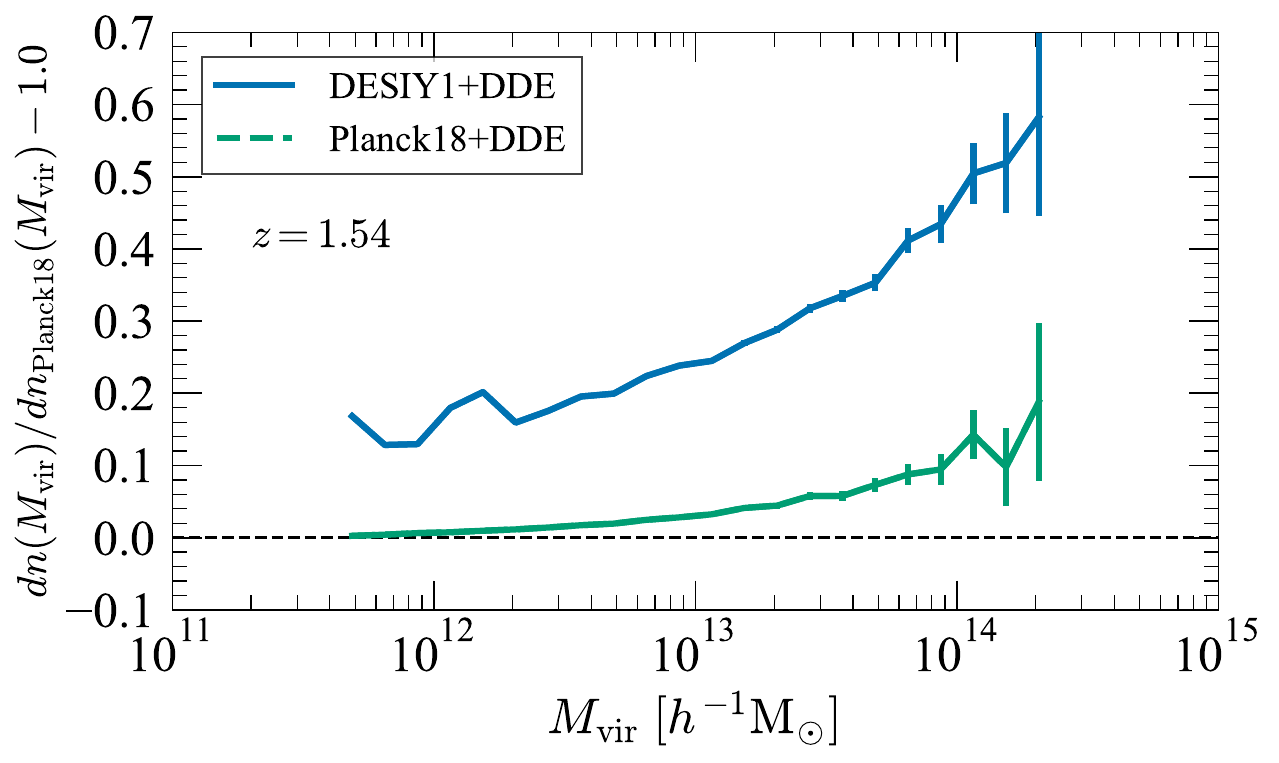}
\includegraphics[width=0.46\textwidth]{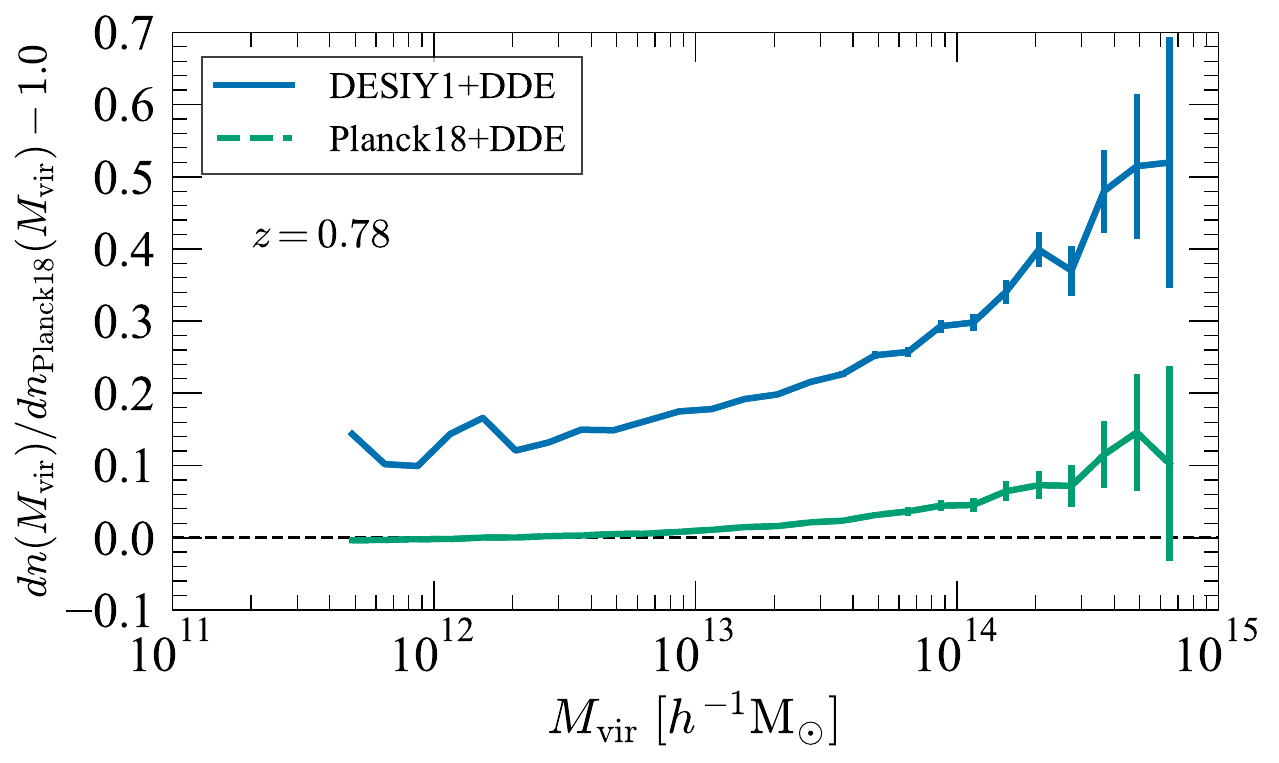}
\includegraphics[width=0.46\textwidth]{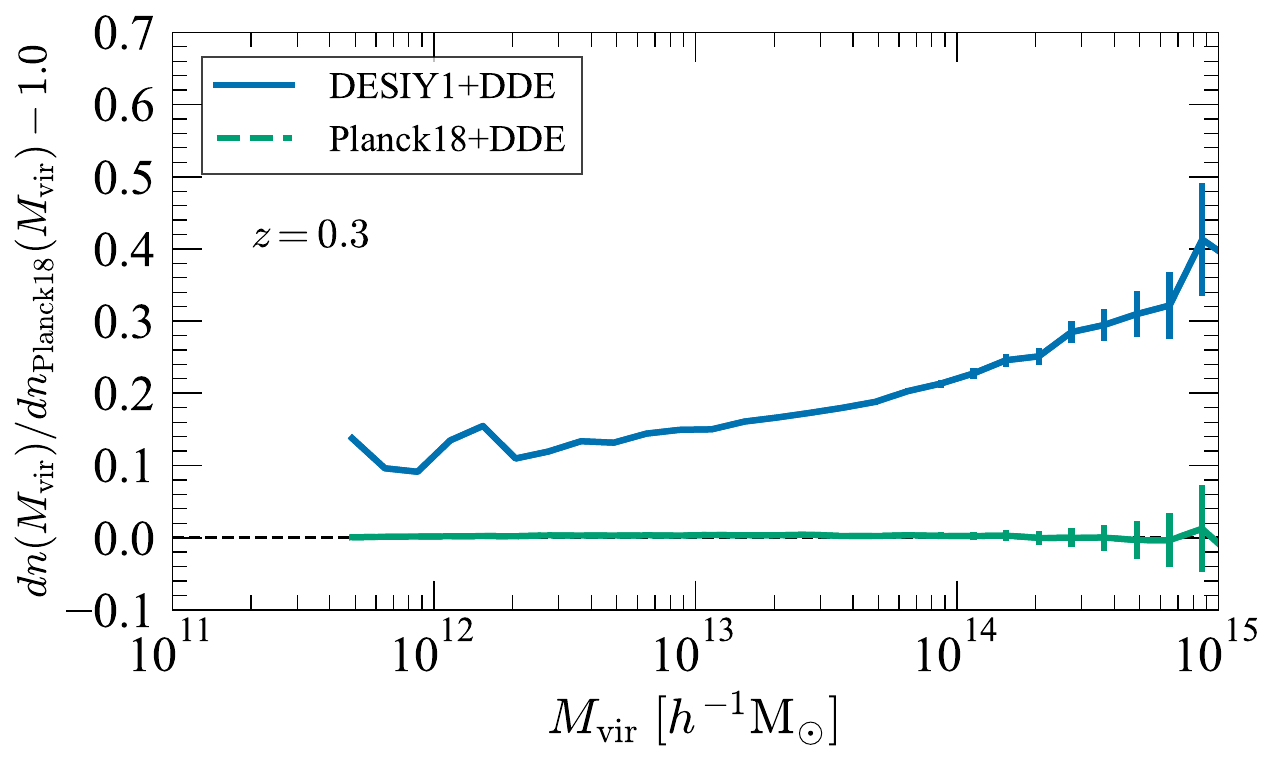}
\caption{Halo mass functions of the \desidde\ and \lcdmdde\ models relative to \lcdm\ simulation at $z=2.03, 1.54, 0.78$, and 0.30.
The relative differences decrease toward lower redshift, and the effect of DDE becomes negligible by $z=0.3$.
The difference in cosmological parameters between the \lcdm\ and \desi-based models has a more significant impact than the effect of DDE.
The error bars correspond to Poisson scatter.
}
\label{fig:mfunc} 
\end{figure*}

\section{Results}\label{sec:result}

We compare the basic clustering properties of dark matter and halos across the three simulation models, focusing on the matter power
spectrum, halo mass function, and the redshift-space two-point
correlation functions and power spectra of halo samples. These
quantities are analyzed at redshifts $z=2.03, 1.54, 0.78, 0.30$, and 0, chosen to match those of the DESI samples of luminous tracers~\cite{DESI2024}.

\subsection{Power spectra}

Figure~\ref{fig:pk} shows the matter power spectra of \desidde\ and
\lcdmdde\ models relative to the \lcdm\ simulation for the range $0.1 < k <
10$\hMpcinv. At redshifts $z\gtrsim 1.54$, the
power spectrum of the \lcdmdde\ model is a few percent higher than that of \lcdm. This difference gradually decreases with
redshifts and becomes negligible (within 1\%) by $z=0.3$. This trend is explained by the evolution of the linear growth factor,
shown in Figure~\ref{fig:model}.  The growth factor in the
\lcdmdde\ model relative to \lcdm\ increases with time, peaking around $z\sim$1, and then gradually
decreases toward $z=0$. A larger (smaller) growth factor enhances (suppresses) 
the power spectrum. However, since the maximum difference in the growth factor
is only about 1\%, the resulting enhancement in power remains at the level of just a few percent.

Comparison of the \desidde\ and \lcdm\ results reveals a more complex situation, primarily due to the differences in the initial matter power spectra,
as shown in Figure~\ref{fig:model}. The BAO pattern around $k \gtrsim
0.1$\hMpcinv\ is visible in the relative power specrtum because the sound
horizon scale is slightly smaller in the \desidde\ model compared to \lcdm~(see
Table~\ref{tab:model}).  This pattern persists in the
relative power at all redshift.  For $k \gtrsim 0.2$\hMpcinv, the initial
power in  \desidde\ is consistently higher than in \lcdm\ by up to
$\sim$10 \%. Additionally, the growth factor is systematically larger in
\desidde\ due to its $\sim$10\% higher value of $\Omega_0$,
which helps counteract the suppressive effect of DDE at $z \gtrsim 0.5$.
As a result, the relative power difference with respect to \lcdm\ is larger
in \desidde\ than in \lcdmdde, except at very small $k$, where
the initial power in \desidde\ is approximately $\sim$15\% lower than in \lcdm.
Overall, the differences in cosmological parameters - particularly $\Omega_0$ - have a more significant impact on the matter power spectrum than the DDE effect alone.

\begin{figure*}
\centering 
\includegraphics[width=0.49\textwidth]{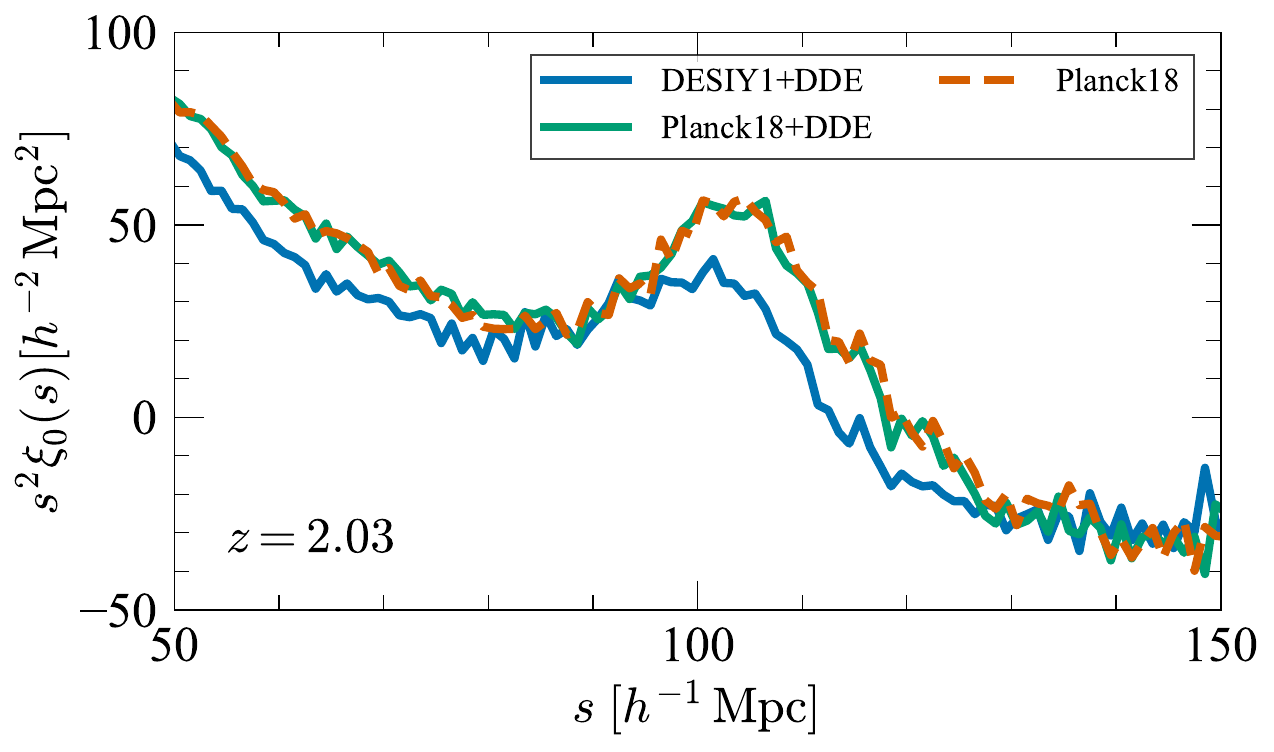}
\includegraphics[width=0.49\textwidth]{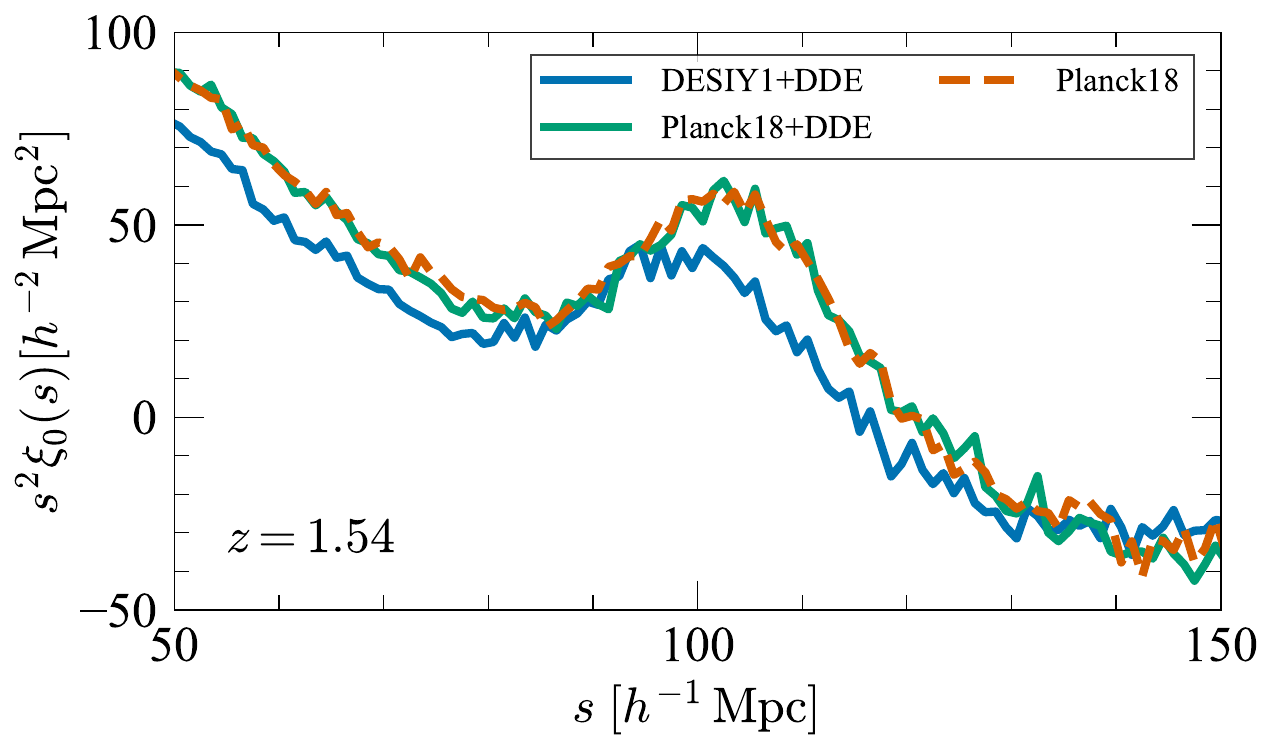}
\includegraphics[width=0.49\textwidth]{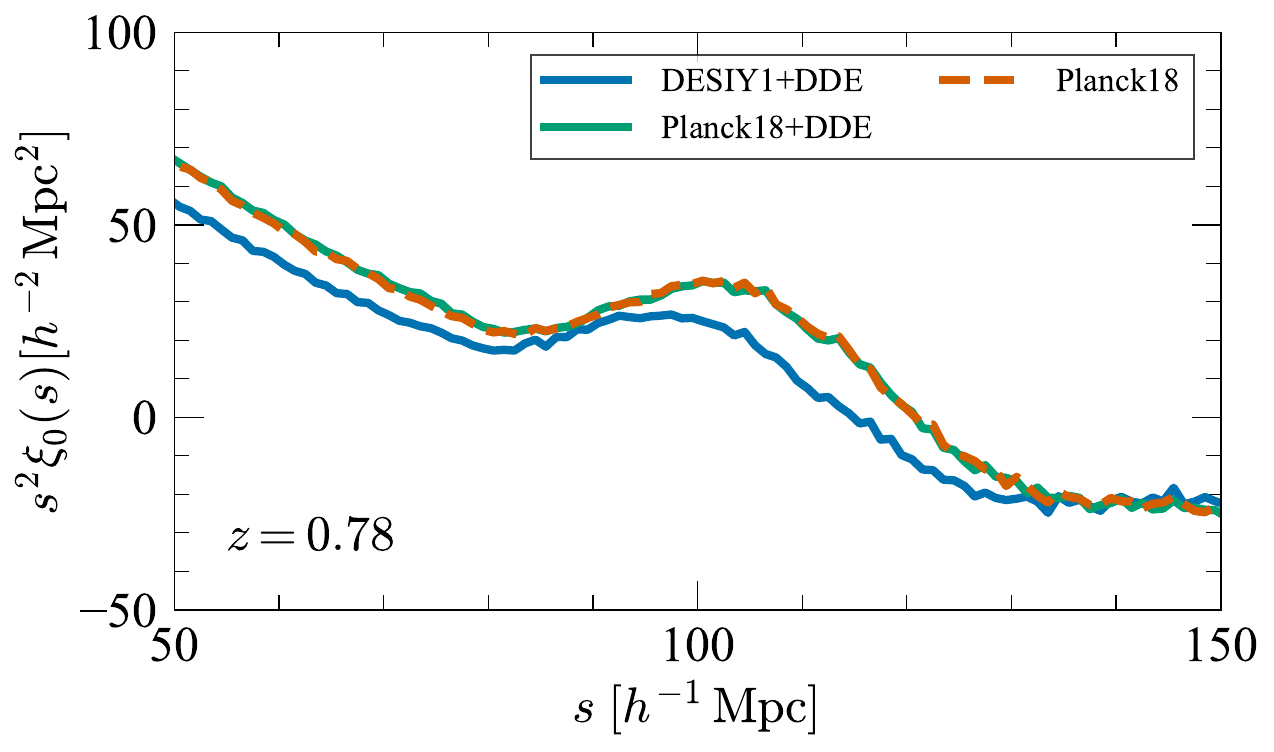}
\includegraphics[width=0.49\textwidth]{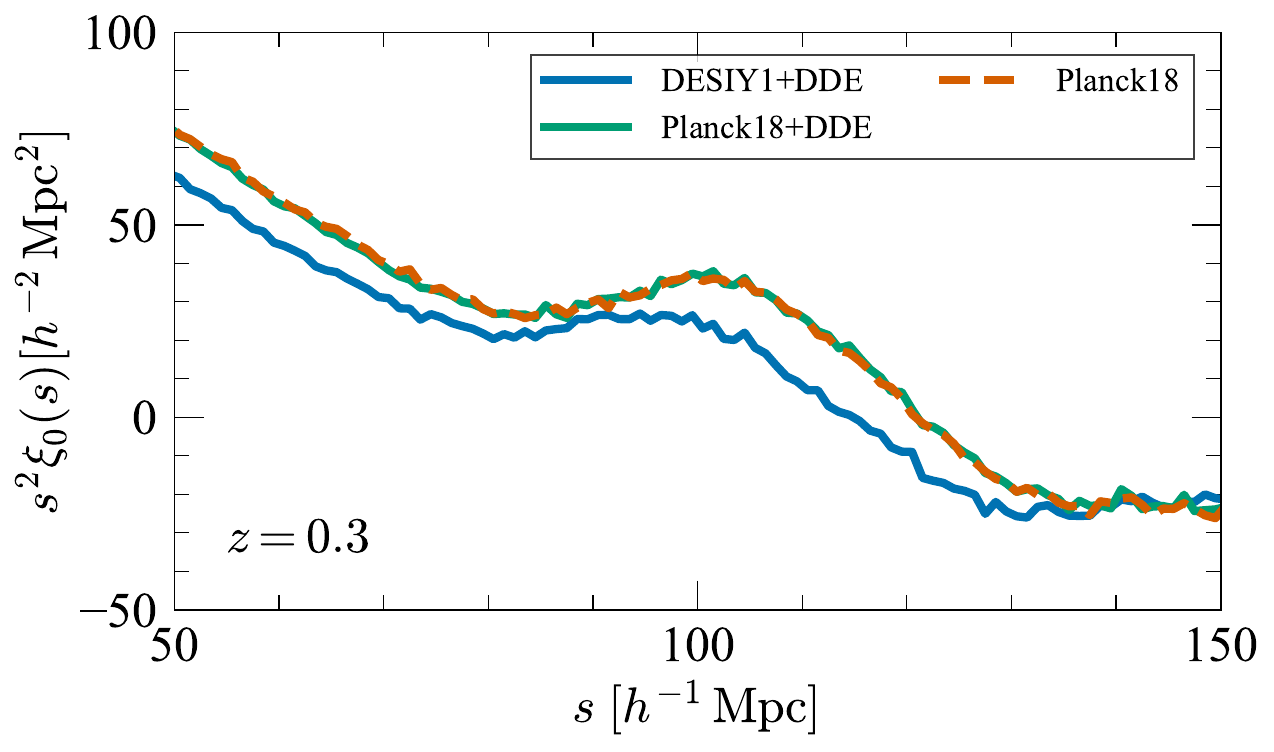}
\caption{
Monopole of redshift-space two-point correlation function, $\xi_0$, scaled by $s^2$, as a function of comoving separation $s$, 
for all three simulations at redshifts  $z=2.03, 1.54, 0.78$, and 0.30, as indicated in each panel. The BAO peak at $s\sim 100$\hMpc\ is clearly visible at all redshifts. In the \desidde\ model, the peak is shifted by approximately $4\%$ toward smaller values of $s$ compared to the \lcdm\ model. This shift is expected due to differences in the sound horizon scales between the models at the drag epoch (see Table~\ref{tab:model}).
}
\label{fig:corrfunc} 
\end{figure*}

\subsection{Halo mass function}

Figure~\ref{fig:mfunc} shows the halo mass functions of the \desidde\ and
\lcdmdde\ simulations relative to the \lcdm\ simulation.  We only show results for halos more massive than $4.0 \times 10^{11}$\hMsun, corresponding to $\sim 40$ particles per halo. 
At all redshift, the \desidde\ model predicts a higher abundance of
massive halos compared to \lcdm, with differences reaching up to $\sim$70 \%. As discussed in  Sec.\ref{sec:DDE}, this is a consequence of the altered shape of the linear power spectrum: the \desidde\ model features a higher amplitude of fluctuations on galactic and cluster scales, which leads to earlier and more efficient formation of halos. The abundance of rare, high-mass halos is particularly sensitive to changes in amplitude, which explains why the excess number of halos in \desidde\ increases with halo mass.

There is a slight excess in the number of halos in the \lcdmdde\ model, but it is dramatically smaller than in the \desidde\ model. Note that \lcdmdde\ and \lcdm\ models share the same cosmological parameters, differing only in the nature of dark energy. Therefore, the difference between these two models arises solely from the variation in the growth rate of fluctuations, not from differences in the initial power spectrum (see the middle panel in Figure~\ref{fig:model}). Since the growth rates differ by only $\sim 1\%$, the resulting change in halo abundance is modest and vanishes by redshift $z=0.3$.

The evolution of the halo mass function exhibits a similar trend to that seen in the
power spectra: the excess relative to \lcdm\ decreases towards lower redshifts for both the \desidde\ and \lcdmdde\ models. In the case of \lcdmdde, the difference  disappears by $z=0.3$.  However, the \desidde\ model still
predicts $\sim$40\% more massive clusters at that redshift, which could serve as
an additional cosmological test.

\subsection{Baryonic Acoustic Oscillations: \lcdm\ vs. \desidde} 

We study the clustering of dark matter halos in redshift space on scales 50$-$150\hMpc, which are relevant for measurements of the BAO feature. Real-space halo positions are displaced into the redshift space along each of the $x$-, $y$- and $z$-axis,
and  the correlation functions are calculated using the \textsc{corrfunc}\footnote{\url{https://github.com/manodeep/Corrfunc}}~\cite{Sinha2019,Sinha2020} code with the Landy-Szalay estimator~\cite{Landy1993}.
The resulting correlation functions along the three axes are averaged and analyzed.

The monopole of the redshift-space two-point correlation function, $\xi_0(s)$, scaled by $s^2$, is shown in Figure~\ref{fig:corrfunc} as a function of the comoving separation $s$. The analysis includes four halo samples drawn from the three cosmological simulations, with number densities (in units of \hMpccube) set to $1.0 \times 10^{-4}$ at $z=2.03$ and 1.54, $4.0 \times 10^{-4}$ at $z=0.78$, and $3.0 \times 10^{-4} $ at $z=0.30$. These $n(z)$ values are chosen to represent the observed number densities of the DESIY1 galaxy and QSO data samples used in BAO measurements~\cite{DESI2024}. 

A distinct BAO peak is clearly visible at $s \approx 100$ \hMpc\ across all samples. The position of this peak is a key cosmological observable, serving as a standard ruler for distance measurements and for inferring the expansion history of the universe~\cite{Eisenstein1998}.

At all redshifts, the BAO peak in the \desidde\ model is shifted by $\sim 4\%$ toward smaller scales (i.e., lower values of $s$) compared to the \lcdm\ model, as expected from differences in their sound horizon scale at the drag epoch (see Table~\ref{fig:model}). In the \desidde\ model, the BAO feature appears sharper and has a slightly reduced amplitude at higher redshifts ($z = 2.03$ and 1.54), where the clustering signal is less affected by nonlinear evolution and redshift-space distortions. In contrast, in the \lcdm\ model, the BAO features become more damped and broader at lower redshifts ($z = 0.78$ and 0.30) due to stronger nonlinear effects. The two-point correlation function of the \lcdmdde\ model is indistinguishable from that of the \lcdm, as both share the same initial BAO pattern in the power spectrum and similar fluctuation amplitudes.

\begin{figure}
\centering 
\includegraphics[width=0.48\textwidth]{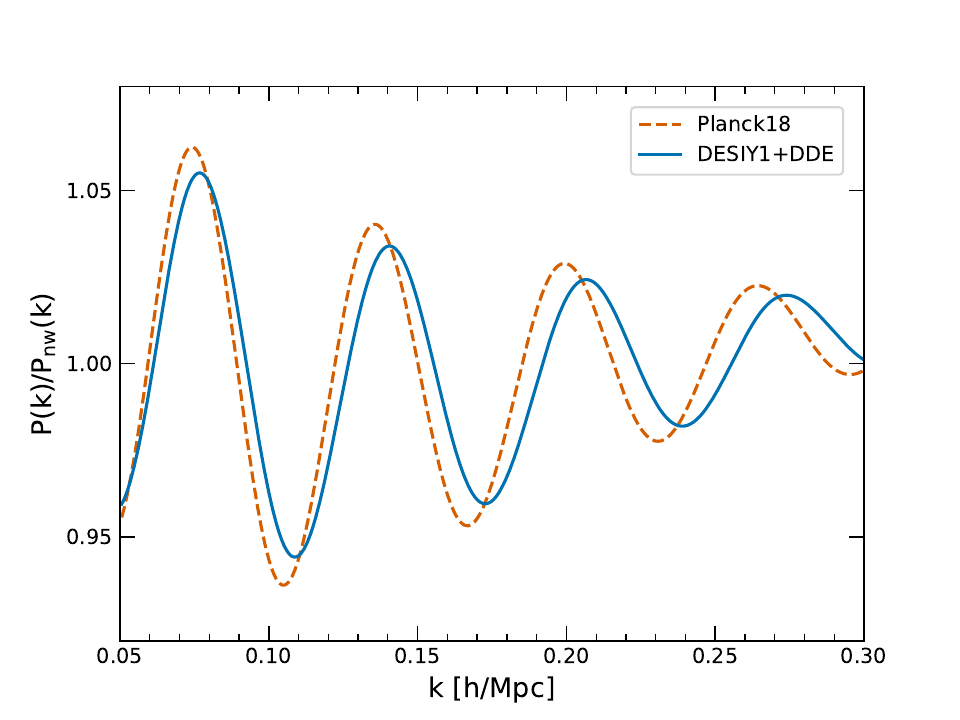}
\caption{Baryonic acoustic oscillations in the linear power spectrum for the \desidde\ (solid curve) and \lcdm\ (dashed curve) cosmologies. The plot shows the deviations of each power spectrum from the corresponding "no-wiggle" reference spectrum - i.e., one without baryonic oscillations - adopted from \citet{Eisenstein1998}. Compared to \lcdm, the BAO peaks in the \desidde\ cosmology are systematically shifted by 3.71\% toward larger wavenumbers, reflecting the difference in the sound horizon scale at the drag epoch.}
\label{fig:PkBAO} 
\end{figure}

Non-linear evolution broadens the BAO features but does not alter the relative shift between the two models, which is set at the last scattering surface, as shown in Figure~\ref{fig:PkBAO}. This figure displays the deviations of each linear power spectrum from the corresponding ‘non-wiggle’ reference spectrum, derived using the fitting formula of \citet{Eisenstein1998} for $P_{\rm nw}(k)$, which excludes the BAO features. 
The separation between the BAO peaks in the power spectrum scales with the isotropic dilation parameter $\alpha$, which accounts for the rescaling of wavenumbers between the two cosmologies. The relationship between the power spectra of the \desidde\ and \lcdm\ models is given by the transformation: 
\begin{equation}
P_{\text{DESI}}(k) = \frac{1}{\alpha^3} P_{\text{PL}}\left(\frac{k}{\alpha}\right).
\label{eq:Pkscaling} 
\end{equation}
The positions of the BAO peaks in $k$-space are systematically shifted according to $k_{\text{DESI}} = \alpha \, k_{\text{PL}}$. 
Thus, the parameter $\alpha$ reflects the scaling of the sound horizon and is defined as the ratio of the sound horizon scales, in units of \hMpc, between the two cosmologies:
\begin{equation}
\alpha = \frac{r_{\rm d}^{\text{PL}}}{r_{\rm d}^{\text{DESI}}}.
\label{eq:alpha} 
\end{equation}
Since \(\alpha > 1\), with \(\alpha = 1.0371\) based on the $r_{\rm d}$ values from Table~\ref{fig:model}, the BAO peak positions in the linear \desidde\ power spectrum are systematically shifted towards larger wavenumbers relative to those in \lcdm\, as shown in Figure~\ref{fig:PkBAO}. 

We measure the BAO shifts in the \desidde\ simulation comoving boxes relative to those from \lcdm\ for the halo samples with number densities and redshifts listed in Table~\ref{tab:BAOfit}. This is done by modeling the non-linear halo power spectra in redshift space, where the acoustic oscillation features of the linear \lcdm\ power spectrum are damped using a Gaussian function. The scale parameter of this Gaussian accounts for the broadening of the BAO features due to non-linear effects. The functional form used follows Eq.~(2) from \citet{Klypin2021} (see references therein). In our model, the BAO shift, quantified by $\alpha_{\rm sim}$, and the damping of the acoustic oscillations are treated as free parameters.

We then fit the power spectra $P(k)$ over the wavenumber range $0.05 < k < 0.3$\hMpcinv. The resulting best-fit values of $\alpha_{\rm sim}$ are provided in Table~\ref{tab:BAOfit} for the halo samples drawn from our \lcdm\ and \desidde\ simulations, listed in the third and fourth columns, respectively, covering redshift from 0 to 1.5. We expect slightly larger BAO shifts in $\alpha_{\rm sim}$, particularly  at higher $k$, with deviations about 0.25\% compared to the sound horizon ratio $\alpha = 1.0371$, due to non-linear effects \citep{Prada2016}. This behavior is indeed observed in the \desidde\ simulation results (\(\alpha^{\rm DESI}_{\rm sim}\)).

\begin{table*}[tbp]
\centering
\caption{
Best-fit values of the BAO shift parameter measured from the  \lcdm\ (\(\alpha^{\rm PL}_{\rm sim}\)) and \desidde\ (\(\alpha^{\rm DESI}_{\rm sim}\)) simulations across different redshifts (\(z\)) for the corresponding halo number densities (\(n\)) in units of \( \text{Mpc}^{-3} h^3 \). The fifth column shows the ratio of the volume-averaged distance between the \desidde\ and \lcdm\ cosmologies, \(D^{\rm DESI}_{\rm V} / D^{\rm PL}_{\rm V}\). The sixth column presents the simulation-based observational dilation parameter, \(\alpha^{\rm DESI}_{\rm obs}\), obtained by converting the comoving simulation box values \(\alpha_{\rm sim}\) using Eq.~\eqref{eq:alphaobs}. The final column lists the theoretical predictions, \(\alpha^{\rm DESI}_{\rm th}\), based on the DESIY1 cosmology.
}
\begin{tabular}{lcccccc}
\hline
$z$ & $n$ & $\alpha^{\rm PL}_{\rm sim}$ & $\alpha^{\rm DESI}_{\rm sim}$ & $D^{\rm DESI}_{\rm V}/D_{\rm V}^{\rm PL}$ & $\alpha^{\rm DESI}_{\rm obs}$ & $\alpha^{\rm DESI}_{\rm th}$\\
\hline
1.54 & $1.0 \times 10^{-4}$  & $0.9957\pm0.0027$  & $1.0374\pm0.0037$  & 0.9970 & $0.9894\pm0.0035$ & 0.98871  \\
0.78 &  $4.0 \times 10^{-4}$  & $1.0012\pm0.0032$ & $1.0390\pm0.0038$  & 0.9826 & $0.9762\pm0.0036$ & 0.97440 \\
0.30 &  $3.0 \times 10^{-4}$  & $0.9994\pm0.0040$ & $1.0395\pm0.0050$  & 0.9838 & $0.9779\pm0.0047$ & 0.97565 \\
0.00 &  $3.0 \times 10^{-4}$ & $0.9964\pm 0.0051$ &$1.0395\pm0.0061$  & 1.0457 & $1.0395\pm0.0061$ & 1.03706  \\
\hline
\end{tabular}
\label{tab:BAOfit}
\end{table*}
\begin{figure}
\centering 
\includegraphics[width=0.52\textwidth]{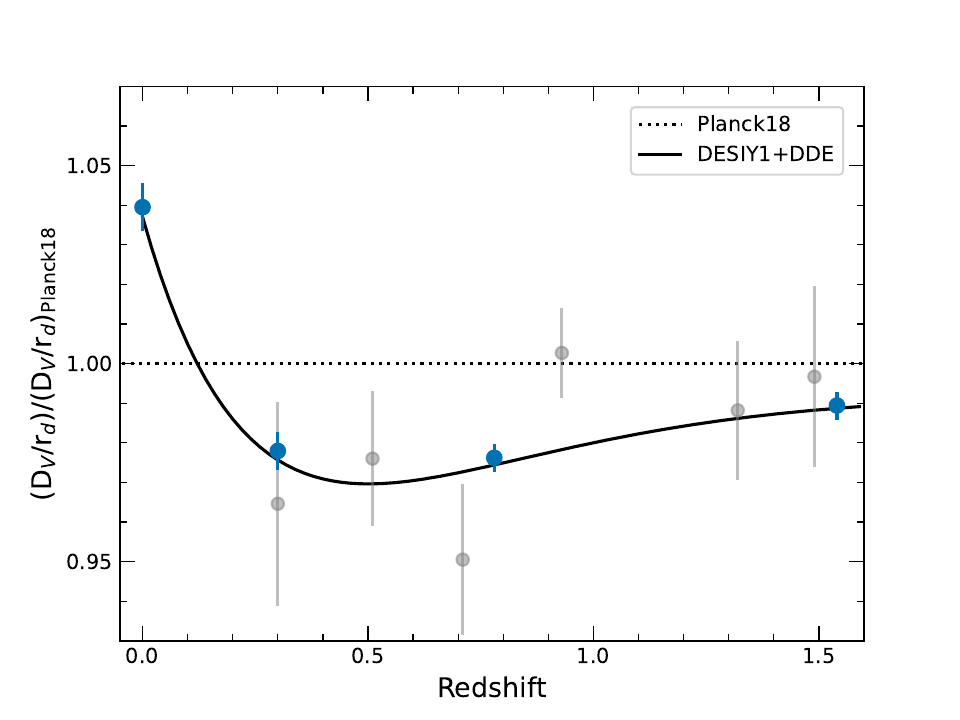}
\caption{Acoustic-scale distance measurements relative to the prediction from the \lcdm\ model. The gray symbols with $1\sigma$ error bars represent the isotropic BAO measurements, $D_{\rm V}(z)/r_{\rm d}$, from DESIY1~\cite{DESI2024}, shown in order of increasing redshift. The curve shows the model prediction from the \desidde\ cosmology, which closely matches the  BAO distance measurements (blue symbols) obtained from the halo samples in our \desidde\ simulation, as listed in Table~\ref{tab:BAOfit}.}
\label{fig:BAOalpha} 
\end{figure}

To covert the BAO shits obtained from the \desidde\ comoving simulation boxes, $\alpha^{\rm DESI}_{\rm sim}$, into the isotropic dilation parameter used in DESI observations, it is necessary to account for cosmology-dependent distances and adopt a fiducial cosmology. Assuming \lcdm\ as the fiducial cosmology, the simulation-based observational dilation parameter $\alpha^{\rm DESI}_{\rm obs}$, is defined as,
\begin{equation}
\alpha_{\text{obs}} = \alpha_{\text{sim}} \frac{h^{\rm PL}}{h^{\rm DESI}} \frac{D_{\rm V}^{\text{DESI}}(z)}{D_{\rm V}^{\text{PL}}(z)},
\label{eq:alphaobs} 
\end{equation}
where $D_V(z) = \left[ (1+z)^2 D_{\rm A}(z)^2 \frac{cz}{H(z)} \right]^{1/3}$ is the volume-averaged (dilation) distance, in units of Mpc~\cite{Eisenstein2005}. Here, \(D_{\rm A}(z)\) is the angular diameter distance (in Mpc), and \(H(z)\) is the Hubble parameter at redshift \(z\) 
(in \kms\ Mpc$^{-1}$). The ratio of the  dimensionless Hubble parameter between the fiducial \lcdm\ and \desidde\ cosmologies accounts for converting the sound horizon ratio, originally expressed in \hMpc, to physical units of Mpc. This correction ensures consistency when comparing the simulation-derived BAO shift parameter, $\alpha_{\text{sim}}$, with observational results. In Table~\ref{tab:BAOfit}, we report the values of $D_{\rm V}^{\text{DESI}}/D_{\rm V}^{\text{PL}}$ and the corresponding observational dilation parameter in the \desidde\ cosmology in the fifth and sixth columns, respectively, computed using Eq.~\eqref{eq:alphaobs}. 

\begin{figure*}
\centering 
\includegraphics[width=0.49\textwidth]{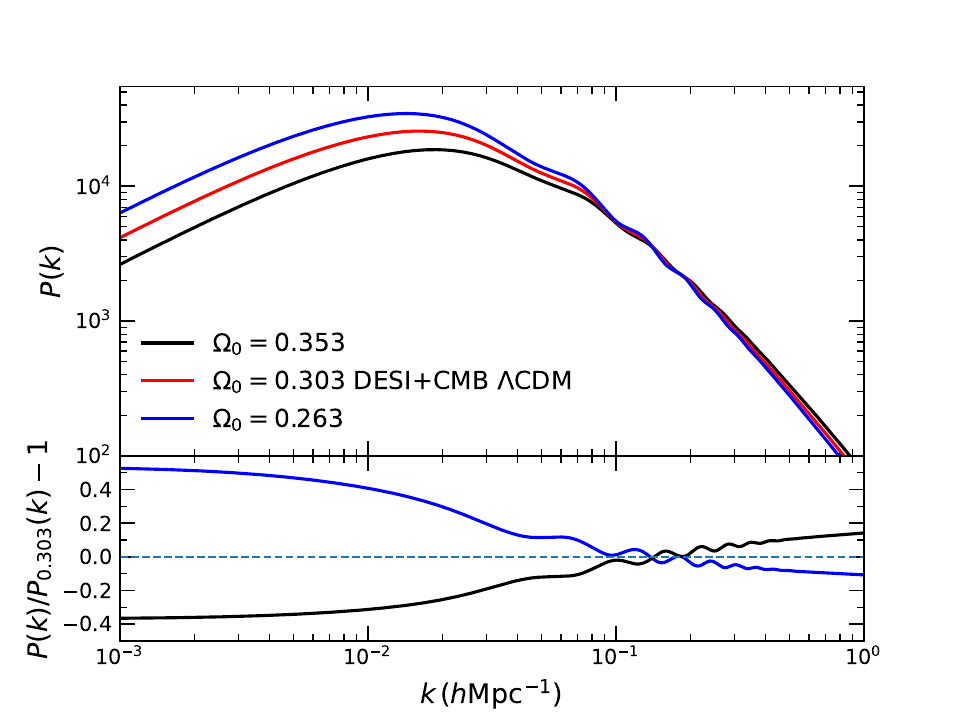}
\includegraphics[width=0.49\textwidth]{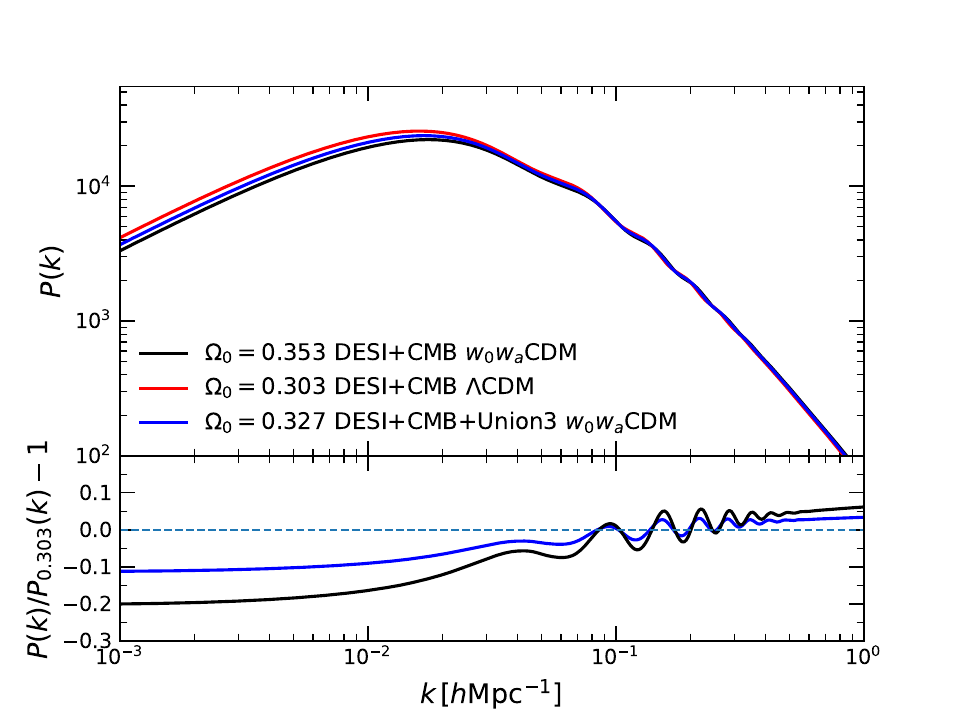}
\caption{Power spectra of dark matter fluctuations for different cosmological models.
All power spectra are normalized to have $\sigma_8=0.810$. The left panel shows models with the same Hubble constant but varying  $\Omega_0$. The model with $\Omega_0=0.303$ corresponds to the DESIY3+CMB+$\Lambda$CDM model \citep{DESIY3}, shown as the red curve. For illustrative purposes, two additional models are included by artificially changing $\Omega_0$ by about 15\%. Note the substantial difference at both small and large scales. The right panel presents models chosen to have the same value of $\Omega_0h^2\approx 0.143$, favored by DESIY3 observations. When this parameter is held fixed, the differences between the models are significantly reduced, although small variations remain.}
\label{fig:Omegas} 
\end{figure*}
The BAO shifts obtained from our \desidde\ simulation, relative to the the predictions from the \lcdm\ cosmology (adopted as the fiducial model), are shown in Figure~\ref{fig:BAOalpha}. The blue symbols represent the simulation results, while the gray symbols correspond to the acoustic-scale distance measurements obtained from the DESIY1 observational data~\cite{DESI2024}. The solid curve shows the model predictions for the \desidde\ cosmology, with the corresponding values at the simulation redshifts listed in the first column  
of Table~\ref{tab:BAOfit}. We find that both the simulation results and the model predictions for the \desidde\ cosmology are in excellent agreement with the observational data. The DDE model predicts a different effective distance scale than $\Lambda$CDM, reflecting modifications to the late-time cosmic expansion rate. At these redshifts, the \lcdm\ model places the BAO peak at a slightly larger separation compared to the \desidde\ model, due to its larger sound horizon. This implies that the fiducial $\Lambda$CDM cosmology predicts a larger volume-averaged distance, $D_V(z)$, at these epochs. 

\section{Discussion}\label{sec:discussion}

How  does time-varying dark energy affects the clustering of dark matter and halos? As we shown above (see, for example, Figures~\ref{fig:pk} and~\ref{fig:corrfunc}), simply modifying the dark energy equation of state -- while keeping other cosmological parameters fixed -- has only a modest impact on clustering. However, once the dark energy is allowed to vary, analysis of DESIY3 observational data yields significantly different best-fit cosmological parameters and power spectra, which in turn lead to noticeable changes in clustering and halo abundances. 
At first glance, this suggests that a wide range of configurations is now possible -- at least at the 10-20 percent level. For instance, DESIY3 results \citep{DESIY3} for the DESI+CMB+$\Lambda$CDM model favor $\Omega_0=0.303$ and $h=0.682$. In contrast, allowing for a time-varying dark energy equation of state (DESI+CMB+$w_0w_a$CDM) gives best-fit values of $\Omega_0=0.353$ and $h=0.636$ -- a significant shift.

The freedom in parameter space is actually more limited than it may initially appear when one considers the power spectrum of fluctuations.  The overall shape of $P(k)$ is tightly constrained by the product $\Omega_0h^2$, which determines the epoch of matter-radiation equality and, consequently,  the location of the turnover scale $k_{\rm max}$ in the power spectrum. Shifting this position leads to significant changes in the shape and amplitude of the power spectrum across a wide range of scales.

\begin{figure}
\centering 
\includegraphics[width=0.45\textwidth]{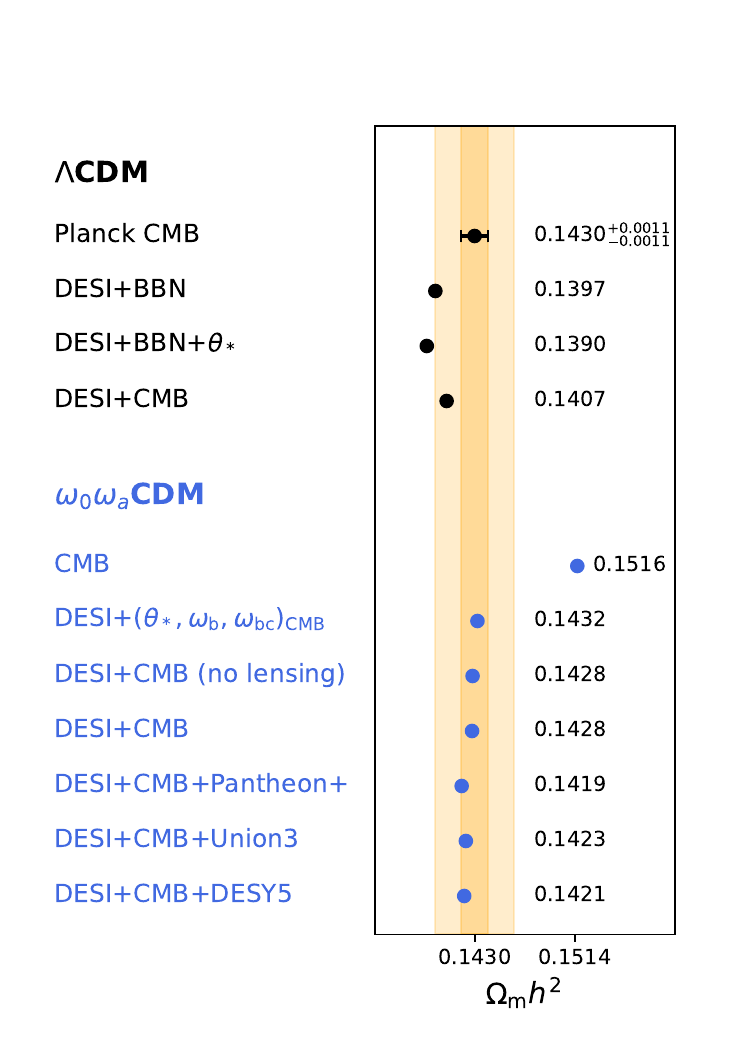}
\caption{Estimates of $\Omega_0h^2$ from various datasets and cosmological models, as analyzed in the DESIY3 paper \citet{DESIY3}. Some estimates are based on the standard $\Lambda$CDM model, while others assume a $w_0w_a$CDM model. Most values lie within a narrow $\sim 1-2\%$ range around $\Omega_0h^2=0.143$. The vertical shaded bends indicate the $1\sigma$ and $3\sigma$ errors derived from Planck CMB data. The only significant outlier corresponds to CMB constraint under the DDE model, which is associated with large intrinsic uncertainties.}
\label{fig:Omh2} 
\end{figure}

The left panel of Figure~\ref{fig:Omegas} illustrates how the power spectrum $P(k)$ responds to changes in $\Omega_0h^2$. In this example, we fix the Hubble constant ($h=0.682$) and keep all other parameters constant ($\sigma_8=0.810, n_{\rm s}=0.96, \Omega_{\rm b}h^2=0.02218$), while varying $\Omega_0$. Setting  $\Omega_0=0.303$ reproduces the power spectrum of DESI+CMB+$\Lambda$CDM model~\citep{DESI2024}, shown as the red curve. For illustrative purposes, we also include two additional models where $\Omega_0$ is artificially increased or decreased by about 15\%. The resulting power spectra are shown as the blue curve (with the same $\Omega_0=0.353$ as in DESI+CMB+$w_0w_a$CDM) and the black curve (with $\Omega_0=0.263$). 

Note how position of the maximum, $k_{\rm max}$, shifts toward larger wavenumbers with increasing  $\Omega_0$. 
This shift is caused by the earlier matter-radiation equality epoch in models with higher $\Omega_0$, which pushes all relevant physical scales to smaller sizes and thus higher $k$ values. We impose the condition that all spectra have the same value of $\sigma_8$. Since the integral defining $\sigma_8$ is dominated by contributions around 
$k\sim 0.1$\hMpcinv, the power spectra are nearly the same in the range $k=0.1-0.3$\hMpcinv. The left panel of Figure~\ref{fig:Omegas} clearly demonstrates that a $\sim 15\%$ change in $\Omega_0h^2$ leads to substantial modifications of the power spectrum: about $40\%$ at $k=0.01$ \hMpcinv\ and $\sim 15\%$ at $k=1$\hMpcinv.

Observations favor a different scenario: while individual estimates of $\Omega_0$ and $h$ can vary as much as $10-20\%$, depending on the observational constraints used, their product $\Omega_0h^2$ remains nearly constant. Figure~\ref{fig:Omh2} shows estimates of $\Omega_0h^2$ from a dozen of different data combinations analyzed in the DESIY3 paper \citep{DESIY3}. Some estimates are based on the standard $\Lambda$CDM model, while others assume DDE models. Remarkably, all estimates fall within a narrow $\sim 1-2\%$ range around $\Omega_0h^2=0.143$. Not surprisingly, this is the value adopted in all our $N$-body simulations.

The right panel of Figure~\ref{fig:Omegas} illustrates what happens to power spectrum when $\Omega_0h^2$ is held fixed. We show results for three models with nearly indentical $\Omega_0h^2=0.143$, but varying matter density: from $\Omega_0=0.303$ for the plain $\Lambda$CDM model (DESI+CDM+$\Lambda$CDM) to $\Omega_0=0.353$ for DESI+CMB+Union3+$w_0w_a$CDM model (see~\cite{DESI2024}). 
When $\Omega_0 h^2$ is fixed, the differences between the power spectra become much smaller, although some differences still remain.

Importantly, DDE models do not significantly modify the inferred value of $\Omega_0h^2$ compared to $\Lambda$CDM. Even across different DDE models -- depending on whether supernova data are included or not -- the estimate of $\Omega_0h^2$ remains robust. This indicates that observational data strongly constrain the shape of the power spectrum, while the impact of dynamical dark energy appears to be secondary.

\section{Summary}\label{sec:summary}

If observations such as those from DESI confirm that a simple dark energy model with a constant equation of state $w$ is insufficient, astronomers will need to consider alternative frameworks~\citep{Park2024,Park2025,Colgain2025,Chakraborty2025,Yin2024,Lee2025}.
To investigate the impact of a time-dependent $w$, we perform three large cosmological $N$-body simulations based on the Chevallier-Polarski-Linder (CPL) parametrization~\cite{Chevallier2001,Linder2003}. The first simulation, labeled \lcdm, serves as a benchmark and follows the the plain $\Lambda$CDM model. The second simulation, \lcdmdde, adopts the CPL model for the dark energy equation of state, while keeping all other parameters identical to those in \lcdm. Comparing these two simulations allows us to isolate the effects of a time-varying equation of state. The third simulation, 
\desidde, includes both a dynamical dark energy component and cosmological parameters constrained by \desi\ data. A comparison between \desidde\ and \lcdmdde\ highlights the  impact of varying the remaining cosmological parameters while holding the dark energy parametrization fixed. The main findings of our study are summarized below.

\begin{itemize}

\item[$^\bullet$] We find that for all statistics we studied -- the matter power spectrum, halo mass function, and halo clustering -- 
the impact of the DDE alone is smaller than the effect of the differences in cosmological parameters between \lcdm\ and \desi.

\item[$^\bullet$] Current observations favor cosmological models with nearly the same value of the product $\Omega_0h^2$~\citep{DESIY3}. Estimates based on both plain $\Lambda$CDM and DDE models consistently fall within a narrow $\sim 1-2\%$ range around $\Omega_0h^2=0.143$.

\item[$^\bullet$] The non-linear matter power spectra show that the \lcdmdde\ model exhibits a small ($\sim 2\%$) enhancement over \lcdm\ at intermediate scales $0.1 < k < 10\,h\,\mathrm{Mpc}^{-1}$ for $z > 1$, driven by differences in the linear growth factor.
By $z = 0.3$, this difference in the power spectra nearly vanishes.
In contrast, \desidde\ exhibits significantly larger deviations relative to the \lcdm\ model. In the linear regime, the power spectrum shows up to 5\% excess at small scales $k \gtrsim 0.2\,h\,\mathrm{Mpc}^{-1}$ and a 15\% suppression at large scales $k < 0.1\,h\,\mathrm{Mpc}^{-1}$. These differences become smaller in the nonlinear regime but remain substantial, reaching 4-6\% at  $k \gtrsim 1\,h\,\mathrm{Mpc}^{-1}$.

\item[$^\bullet$] The halo mass function reflects these trends, with \desidde\ predicting up to 70\% more massive halos at $z = 2$, and sustaining a 40\% excess at $z = 0.3$. 

\item[$^\bullet$] Clustering analysis reveals a 3.71\% shift of the BAO peak towards smaller scales in the \desidde\ model, resulting from the reduced sound horizon scale compared to \lcdm. Measurements of the $\alpha$ dilation parameter,  using halo samples with DESI-like tracer number densities across redshifts $0 < z < 1.5$, yield values consistent with the \desidde-to-\lcdm\  sound horizon ratio. After accounting for cosmology-dependent distances, the simulation-based observational dilation parameter closely matches DESI Y1 observations. 
\end{itemize}

\section*{acknowledgments}

This work has been supported by IAAR Research Support Program in Chiba
University Japan, MEXT/JSPS KAKENHI (Grant Number JP19KK0344 and
JP23H04002), MEXT as ``Program for Promoting Researches on the
Supercomputer Fugaku'' (JPMXP1020230406), and
JICFuS.  Numerical computations were carried out on the supercomputer
Fugaku provided by the RIKEN Center for Computational Science (Project
ID: hp240184).
FP and AK acknowledge support from the Spanish MICINN PID2021-126086NB-I00 funding grant. The data release used the the Skies \& Universes database, hosted at the skun@IAA computer managed by the IAA-CSIC in Spain, through MICINN EU-Feder grant EQC2018-004366-P.

\section*{Data availability}

The data that support the findings of this article are openly available~\cite{Ishiyama2025data}.


\end{document}